# Antiferromagnetic Coupling between Surface and Bulk Magnetization and Anomalous Magnetic Transport in Electro-deposited Co Film

Surendra Singh,[1,*] C. L. Prajapat,[2] D. Bhattacharya,[1] S. K. Ghosh,[3] M. R. Gonal[4] and S. Basu[1]

[1]Solid State Physics Division, Bhabha Atomic Research Centre, Mumbai, 400085 India
[2]Technical Physics Division, Bhabha Atomic Research Centre, Mumbai, 400085 India
[3]Material Processing Division, Bhabha Atomic Research Centre, Mumbai, 400085 India
[4]Glass and Advanced Material Division, Bhabha Atomic Research Centre, Mumbai, 400085 India

*Email: surendra@barc.gov.in

**Abstract:**

We report an interesting magnetic behavior of a Co film (thickness ~ 350 Å) grown on Si/Ti/Cu buffer layer by electro-deposition (ED) technique. Using depth sensitive X-ray reflectivity and polarized neutron reflectivity (PNR) we observed two layer structures for the Co film grown by ED with a surface layer (thickness ~ 100 Å) of reduced density (~ 68% of bulk) compared to rest of the Co film (thickness ~ 250 Å). The two layer structure is consistent with the histogram profile obtained from atomic force microscope (AFM) of the film. Interestingly, using PNR, we found that the magnetization in the surface Co layer is inversely (antiferomagnetically) coupled (negative magnetization for surface Co layer) with the rest of the Co layer for the ED grown film. While we compare PNR result for a Co film of similar layered structure grown by sputtering, the film showed a uniform magnetization as expected. We also show that the depth dependent unusual magnetic behavior of ED grown Co film may be responsible for anomalous anisotropic magnetoresistance observed in low field in this film as compared to the Co film grown by sputtering. Combining X-ray scattering, AFM, superconducting quantum interface device magnetometry (SQUID), PNR and magneto-transport measurements we attempted to correlate and compare the structural, magnetic and morphological properties with magneto-transport of Co films grown by ED and sputtering. The study indicates that the interesting surface magnetic property and magneto-transport property of the ED film is caused by its unique surface morphology.

**Introduction:**

The interplay between morphology, electrical transport and magnetic field in various thin films and multilayer systems of inorganic, ferromagnetic materials, have been demonstrated by the observations of different phenomena e.g. anisotropic magnetoresistance (AMR) [1], tunneling magnetoresistance (TMR) [2], and giant magnetoresistance (GMR) [3]. These phenomena have significant contributions both in understanding the fundamental physical principles of spin transport [4] and in practical applications [5]. AMR plays an important role in the magneto-transport of a single ferromagnetic layer. Ferromagnetic thin films (e.g. Co, Ni, Fe, NiFe) are widely implemented in sensors and computer hard disks because of their AMR property. However the role of reduced site coordination and symmetry at the surface, morphology of the film, which depends on the deposition techniques, may drastically change the AMR properties of the films.

Cobalt serves also as a model system for the macroscopic magnetic response; because the low to moderate crystal anisotropy allows the effects of size, shape, internal crystal structure, and surface anisotropy to be observed in a single system [6, 7]. Recently there have been many experimental and theoretical studies on magnetic properties of Co nanoparticles grown on metallic surfaces of Cu [8-10], Pt [11-13], Ag [14]. Magnetic and magnetotransport properties of Co/Cu multilayer grown by different deposition techniques has also been investigated extensively [15-18]. Abnormal magnetic behavior (negative remanence magnetization) is also observed in a randomly distributed Co nano particle system on Si substrate [19].

Magnetization reversal processes in ultrathin films are also of fundamental importance in both basic research and for technological applications [20, 21]. Gruyters et al. [22], found difference in surface and bulk magnetization during magnetization reversal in Co thin film. The study suggested that this reversal behavior of surface magnetization depend upon the surface chemical composition. Recently Bi *et al.,*[23] observed a voltage controlled magnetization of Co film which was attributed to formation of voltage-induced *reversible oxidation* of the Co layer at the interface. The role of reduced site coordination and symmetry at the surface, at interfaces, at steps, or at islands are some of the key elements for the precise control of the domain structure and of the magnetization direction. These are highly influenced by the growth techniques of the film.

Electro-deposition (ED) is a widely used technique for producing alloys and multilayers by controlling the film composition and thicknesses at an atomic scale by regulating the pulse amplitude and width [24, 25]. The existence of the electric field of the order of $10^7$ V/cm between the electrode and ions as well as the charged nature of the particles arriving at the surface during the growth process are possible factors expected to cause different properties of the electrodeposited film from those being produced from the vapor phase (sputtering). Magnetic thin films grown by ED have shown different morphology [26] and exceptionally high magnetic anisotropy of the perimeter atoms of 2D nanoclusters formed during the growth [27].

Here we report the experimental evidence of the presence of a surface Co magnetic layer (thickness ~ 100 Å) showing negative magnetization with respect to the rest of magnetic Co layer for a Co film grown on Si (substrate)/Ti/Cu surface by ED. We have also compared the magneto-transport properties and their correlation with the depth dependent structure, magnetic and morphological properties of Co films grown by ED and sputtering. Structurally the surface morphology of the ED film is quite different from the sputtered film. We observed anomalous AMR for ED deposited Co film whereas the Co film deposited by sputtering show negligible AMR. Detailed depth dependent structure and magnetic properties of the Co films were obtained from X-ray reflectivity (XRR) and polarized neutron reflectivity (PNR) data. The crystallographic structure, macroscopic magnetization and morphological characterization of Co films were carried out using x-ray diffraction (XRD), superconducting quantum interface device (SQUID) and atomic force microscopy (AFM), diffuse (off-specular) x-ray reflectivity (DXRR) techniques. The depth dependent unusual magnetic behavior and anomalous MR of ED grown Co film may possibly have been induced by the unique surface morphology of the ED grown film.

**Experimental**

Two Co thin film samples of thickness ~ 350 Å each were grown on Si substrate by ED and magnetron sputtering techniques. The Co film grown by ED, henceforth named as sample S1, was electrodeposited on 250 Å thick Cu seed layer on a buffer layer of Ti of thickness 150 Å. Both buffer and seed layer are the requirement of getting good adhesion and quality of Co layer by ED. The nominal structure of sample S1 can be represented as Si(substrate)/Ti (150 Å)/Cu(250 Å)/Co(350 Å). First, both seed and buffer layers were grown by magnetron sputtering at a base vacuum ~$5\times10^{-7}$ Torr, then the Co film was electrodeposited on the Si/Ti/Cu surface using an electrolyte containing 1.5M $CoSO_4 \cdot 7H_2O$ and keeping temperature and pH constant at 298 K and 2.7, respectively, during the deposition. ED was carried out potentiostatically at -1.4 V *vs.* SCE (saturated calomel electrode) with a deposition rate of 3.1 nm.s$^{-1}$ as optimized by electrochemical quartz crystal microbalance (EQCM). For comparing the structure, magnetic and magnetotransport properties we also deposited another Co film by magnetron sputtering on similarly grown (by sputtering) Si/Ti/Cu surface, henceforth named as sample S2. As-deposited structure S2 can be represented as Si (substrate)/Ti (150 Å)/Cu(250

Å)/Co(350 Å). The exact thickness, roughness and morphology of the layers in these samples were obtained by reflectometry techniques discussed later.

We have used Cu K$_\alpha$ (wavelength = 1.54 Å) radiation for XRR and XRD measurements. PNR measurements were carried out at Dhruva, BARC, Mumbai [28], using neutrons of wavelength 2.5 Å. An in-plane magnetic field of 1.7 kOe was applied on the samples during PNR measurements. Macroscopic magnetization characterization of the films was carried out using SQUID magnetometer (Quantum Design, model MPMS5). Transport measurements were carried out using four-probe technique with varying the magnetic field in magnetoresistance (MR) measurement system (OXFORD, model TelstronPT). The morphologies of Co surfaces of S1 and S2 were investigated using an NT-NDT's Solver P-47 H multimode atomic force microscope (AFM) instrument. NSG10_DLC super sharp DLC tip grown on silicon with curvature 1-3 nanometer has been utilized in semi-contact mode. In AFM operation cantilever's resonant frequency & force constants were 213 kHz and 10 N/m respectively. We also carried out DXRR measurements to quantify the surface/interface morphology of our samples in terms of the parameters of a height-height correlation function.

Specular (angle of incidence = angle of reflection) XRR and PNR are nondestructive techniques from which the depth dependent structure of the sample with nanometer resolution averaged over the lateral dimensions of the sample (typically 100 mm$^2$) can be inferred [29-36]. The specular reflectivity is related to the square of the Fourier transform of the depth dependent (z) scattering length density (SLD) profile $\varphi(z)$ (normal to the film surface or along the z-direction) [29-31]. For XRR, $\varphi_x(z)$ is proportional to electron density whereas for PNR, $\varphi(z)$ consists of nuclear and magnetic SLDs such that $\varphi^\pm(z) = \varphi_n(z) \pm CM(z)$, where $C = 2.9109 \times 10^{-9}$ Å$^{-2}$ m/kA, and $M(z)$ is the magnetization (kA/m) depth profile [29-31]. The sign +(-) is determined by the condition when the neutron beam polarization is parallel (opposite) to the applied field and corresponds to reflectivities $R^\pm$.

**Results and Discussion**

Fig. 1 (a) and (b) show the XRD measurements from S1 and S2, respectively. Both the films show fcc crystalline structures for Cu layer with textured growth along (111). It is noted that XRD from S1 also shows a small but wide peak corresponding to fcc Co (111). The different Bragg reflections corresponding to different planes of fcc Cu and Co are indexed in Fig. 1 (a). While both the samples S1 and S2 show strong texture and similar crystallinity, the most prominent difference lies in their surface morphologies, discussed later.

Fig. 2 (a) shows the SQUID magnetization, $M(H)$, measurements from S1 at room temperature along two perpendicular in-plane directions as indicated in the inset. Hysteresis curve from S1 suggest a soft ferromagnet with a coercive field ($H_c$) of ~ 45 Oe and saturation magnetization of ~ 820 kA/m. The ratio of remanence magnetization ($M_{rem}$) and saturation magnetization ($M_{sat}$) for the sample is ~96%. Fig. 2 (b) shows the $M(H)$ measurements for S2 at room temperature along two perpendicular in-plane directions (x and y –axis). S2 shows relatively lower saturation magnetization of ~ 618 kA/m with a coercive field of ~ 36 Oe. The ratio $M_{rem}/M_{sat}$ is ~53% for S2. It is evident from Fig. 2 (a) and (b) that hysteresis curves for two samples indicate that S1 has a higher saturation magnetic moment with a broader hysteresis loop compared to S2.

MR, defined as $[(\rho(H) − \rho(H_s))/ \rho(H_s)]$, where $\rho(H)$ and $\rho(H_s)$ are resistance as a function of magnetic field ($H$) and resistance at saturation field ($H_s$), respectively. MR measurements from S1 as a function of magnetic field along three in-plane direction are depicted in Fig. 2 (c). The applied magnetic field ($H$) was oriented in-plane with respect to the direction of current ($I$) as shown schematically in the inset of Fig. 2 (c). The low-field MR is positive for in-plane magnetic fields (both $I$ and $H$ along same direction, longitudinal geometry) and negative for perpendicular applied fields ($H$ is at 90 degree from $I$ but in the plane, transverse geometry). Hysteresis is also evident, which correlates well with magnetization hysteresis loops (Fig. 2(a)). The magnetotransport properties of S1 exhibit AMR behavior but the variation of low field (-100 Oe < H < 100 Oe) MR is opposite (inverse transverse and longitudinal MR) to the MR behavior observed for Co based thin films and multilayers reported earlier [1, 37-40]. The anomalous AMR behaviour of S1 might have resulted due to different magnetic and morphological properties of Film grown by ED as discussed later. We also observed much small MR with finite hysteresis when $H$ is applied at an angle ($\varphi$) of $45^0$ from the direction of $I$. Fig. 2 (d) shows the MR measurement from S2 as a function of magnetic field in longitudinal geometry. We also observed similar MR results from S2 in transverse geometry. We obtained very small MR and without any peak in MR for magnetic field around $H_c$ for S2. Thus MR from S2 show completely different variation with respect to applied field as compared to that of S1.

In order to understand the origin of different in-plane macroscopic magnetic and transport properties of these Co films we have studied the depth dependent structure and magnetic properties of the films using XRR and PNR techniques [28-36]. XRR and PNR involve measurement of the x-ray/neutron radiation reflected from a sample (inset of Fig. 3) as a function of wave vector transfer $Q$ (i.e., the difference between the outgoing and incoming wave vectors). In case of specular reflectivity (angle of incidence, $\theta_i$ = angle of reflection, $\theta_f$), $Q = Q_z$ [$= \frac{2\pi}{\lambda}\left(sin(\theta_i) + sin(\theta_f)\right)$, where $\lambda$ is wavelength of x-ray/neutron] and $Q_x$ [$= \frac{2\pi}{\lambda}\left(cos(\theta_i) - cos(\theta_f)\right)$] = 0, thus we get depth dependent profile of structure and magnetism of the sample [33, 34]. Whereas in case of diffuse (off-specular) reflectivity ($\theta_i \neq \theta_f$), we have reflectivity as a function of $Q_x$ (at fixed $Q_z$) and provide in-plane morphology of the sample [33, 34, 36, 41-42].

Fig. 3 (a) and (b) show the specular XRR as a function of $Q_z/Q_c$, where $Q_c = 4\pi sin\theta_c/\lambda$ and $\theta_c$ is critical angle of incidence (below which reflectivity is unity) for S1 and S2, respectively. The electron SLD (ESLD) was inferred from the XRR data by fitting a model $\varphi(z)$ whose reflectivity best fits the XRR data (solid lines in Fig. 3). A model consisted of layers representing regions with different ESLDs. The parameters of the model included layer thickness, interface (or surface) roughness and ESLD. The reflectivity was calculated using the dynamical formalism of Parratt [35], and parameters of the model were adjusted to minimize the value of weighted measure of goodness of fit, $\chi^2$ [43]. The inset of Fig. 3 (a) and (b) shows the ESLD depth profile of S1 and S2, respectively, which best fitted (solid lines in Fig. 3) the XRR data. We observed a low density Co layer of thickness ~ 100 Å at the surface for the S1 (grown by ED). We also observed higher roughness for ED grown interfaces as compared to that grown by sputtering. We obtained lower ESLD (~ 87% of bulk) for Co layer of S2 (grown by sputtering). The parameters obtained from XRR data for S1 and S2 are given in Table 1.

The PNR data at room temperature from S1 and S2 are shown in Fig. 4. PNR data were measured up to lower $Q/Q_c$ (~ 3) values as compared to XRR data (~ 6). Thus using the structural parameters (thickness of layers, interface roughness and number density of each layer) obtained from XRR data as a guide, we have fitted the PNR data from the samples. Fig. 4(a) and (b) show the spin dependent reflectivity data (closed and open circles) and corresponding fits (solid lines) to the data from S1 and S2, respectively. The nuclear scattering length density (NSLD) and magnetization (M) depth profile of the samples which best fitted PNR data are shown in Fig. 4 (e)-(h). Fig. 4 (c) and (d) show the spin asymmetry (SA) data from S1 and S2, respectively, defined as $(R^+ - R^-)/(R^+ + R^-)$, which highlights the spin dependence of the reflectivity. The amplitude and period of the oscillatory variation in SA are related to the magnetization contrast across interfaces between magnetic/non-magnetic layers and the total thickness of the film respectively [30]. The structural parameters extracted from the PNR data of S1 and S2 are given in Table 2, which are consistent with XRR results (Table 1).

The important finding in present study, as revealed by combined results of XRR and PNR, is formation of a low density (~ 68% of bulk) Co layer for ED grown Co film (S1), which is antiferomagnetically coupled (negative magnetization as compared to rest of Co layer) with rest of Co layer. We obtained an average magnetization of ~ -500 kA/m for low density Co layer and ~1430 kA/m (bulk value of magnetization for Co is ~ 1440 kA/m ) for rest of Co layer (Fig. 4 (g)) of S1. The thickness averaged magnetization of the entire Co layer in S1 obtained by PNR is 801±20 kA/m, which is consistent with the magnetization of ~820 kA/m obtained from macroscopic SQUID magnetometer, but lower then bulk magnetic moment of Co. We also obtained a reduced magnetization for Co layer in S2 (Fig. 4 (h)) as compared to bulk value. The thickness averaged magnetization of Co layer in S2 as obtained from PNR measurement was 635±15 kA/m which is also consistent with the saturation magnetization (~ 620 kA/m) of S2 as obtained from SQUID. Reduced saturation magnetization of polycrystalline Co film was also observed earlier and attributed to adjacent layers material and the nonlocal environment [44].

Further to investigate the nature of low density Co layer on the surface of S1 we have again used the intrinsic properties of XRR and PNR, which distinctly reveal the composition of the layer [45, 46]. It is noted that XRR results of S1 showed a Co layer (thickness ~ 100 Å) with lower ESLD (~$4.6 \times 10^{-5}$ Å$^{-2}$) at the surface, which could be CoO as the observed ESLD is very close to that of bulk value for CoO (~$4.76 \times 10^{-5}$ Å$^{-2}$). However the NSLD for this layer (surface) is ~$1.72 \times 10^{-6}$ Å$^{-2}$ as obtained from PNR data, which neglects this possibility (CoO), since NSLD for bulk CoO (~$4.29 \times 10^{-6}$ Å$^{-2}$) is much larger than the NSLD for Co (~ $2.27 \times 10^{-6}$ Å$^{-2}$). This has been demonstrated in Fig. 5. Assuming the top layer as CoO with zero average magnetization (since CoO is antiferromagnetic at room temperature) we tried to fit the SA (or PNR) data from S1. Fig. 5 (a) and (b) show the SA data and corresponding NSLD and magnetization depth profile. It is evident from Fig. 5 that considering a top layer of thickness ~ 100 Å as CoO resulted into poor fit to the SA (PNR) data. Thus the results indicate that the top layer near the surface is a low density Co layer. However a very small oxide layer of thickness (< 10 Å) at the surface cannot be ruled out as the presented reflectivity data is not sensitive to resolve such thickness. Formation of low density of Co layer might have resulted due to different nucleation process of the Co layer deposited by ED.

In order to verify the uniqueness of obtained magnetization profile, a negative magnetization on the surface of S1, from PNR, we have also attempted several magnetization depth profiles to fit the PNR data from S1. Three models assuming different magnetization depth profiles with fixed NSLD (shown in Fig. 4 (e)), as well as corresponding fit to the SA (PNR) data are shown in Fig. 6. The models are (i) uniform magnetization within whole Co layer; reduced magnetization for top low density Co layer but the magnetization is (ii) parallel and (iii) anti-parallel to the magnetization of rest of the Co layer. The magnetization models (i), (ii) and (iii) are shown in Fig. 6 (d), (e) and (f), respectively, and the corresponding fits to the SA data are shown in Fig. 6 (a), (b) and (c), respectively. It is evident from Fig. 6 that a reduced magnetization of top low density Co layer with magnetization oriented to the opposite of the magnetization of high density (bulk density) Co layer gave best fit to the PNR data from S1.

From the depth dependent structure and magnetic properties as extracted using XRR and PNR, we found a specific magnetic behavior of Co film of S1. The surface layer (~ 100 Å Co layer with reduced density) and bulk of Co film are coupled antiferromagnetically even at a higher magnetic field (~ 1.7 kOe). Also such a magnetic behavior may be responsible for observed anomalous MR, because SQUID magnetometer did not show any change in $M$(H) hysteresis curve of the Co films along two perpendicular in-plane direction. Whereas a uniform but reduced magnetization for Co film of S2 was observed. Strain and morphology induced large magnetic anisotropy are two important factors which have contributed strongly for influencing the magnetization and magnetic transport properties of thin films and nanostructures [27, 47-52]. Strain-induced magnetization reorientation and very large magnetic anisotropy of Co based thin films have been reported earlier [47-49]. In addition negative magnetization for Co based oxide systems have also been reported earlier [50] which is attributed to negative exchange coupling among ferromagnetic sublattices due to magneto crystalline anisotropy. Uniaxial anisotropy and growth condition were also attributed for the observation of negative differential magnetization in ultrathin Fe film [51]. An exceptionally high magnetic anisotropy of the perimeter atoms of 2D nanoclusters formed during growth of Co film by ED was also seen earlier [27], which is attributed to distinct growth and subsquent morphology of the film. The huge magnetic anisotropy behavior of the islands' edges [51, 52] due to distinct morphology also tuned the magnetic properties [52] of two dimensional nanostructures.

In order to qualitatively see the effect of strain for unusual magnetic and AMR properties we compared and estimated the strain for Co film of S1 and S2 by revisiting the XRD data from these samples. We observed Bragg peaks corresponding (111) reflections from both Cu ($2\theta \sim 43.33^0$) and Co ($2\theta \sim 44.25^0$) for S1, whereas S2 showed a single peak at higher angle ($2\theta \sim 43.45^0$). Fig. 7 (a) and (b) show the XRD data around (111) reflections from sample S1 and S2, respectively. We have fitted two Gaussian for (111) reflection from sample S1 [Fig. 7(a)] which suggested peaks at $2\theta \sim 43.33^0$ (bulk Cu (111) reflection, $d_{111}$ = 2.086 Å) and $44.25^0$ (Bulk Co (111) reflection, $d_{111}$ = 2.044 Å) with a FWHM of $0.5377^0$ and $1.0867^0$, respectively. Using Debye-Scherrer formula [53]: [$= \frac{0.9 \lambda}{\beta \cos\theta_b}$; where $\lambda$, $\beta$ and $\theta_b$ are wavelength of x-ray, line broadening and Bragg angle], we estimated crystallite sizes of Cu and Co in S1 to be ~17 nm and ~8 nm, respectively. The (111) reflection from S2 [Fig. 7(b)] fitted well with single Gaussian around $2\theta \sim 43.45^0$ ($d_{111}$ = 2.080 Å) with a FWHM of $0.49261^0$, suggesting a crystallite size of ~ 19 nm. The results suggest that Cu atoms in S2 are under small compressive strain of ~ 0.3%, whereas Co atoms in S2 are under large tensile strain of ~ 1.77%. This may be one of the reasons for obtaining lower density (~ 90%) for whole Co layer for sputtered sample (S2) and hence for reduced magnetization of the Co film in this sample. However the Co film in S1 is relaxed and suggests negligible effect of strain on unusual magnetic and magnetotransport properties of S1.

This prompted investigating the comparative morphological properties of surface and different interfaces of S1 and S2. We have used AFM and DXRR measurements for quantifying the morphologies. Fig. 8 (a) and (b) show the AFM images, with a scan size of 2 μm × 2 μm, on the surfaces of S1 and S2, respectively. The AFM image of S1 indicates an island growth (clustering) of the film with a cluster size varying from few hundred nanometers to submicron length scale. Also the cluster (grain) are dispersed and non-uniformly distributed, resulting in a porous surface with hills separated by large valley like morphology. Whereas the AFM image of S2 indicates that small Co islands uniformly cover the surface, resulting in a smooth or a surface of lower amplitude of undulation as evident from the AFM images. The corresponding topographical histograms of the samples are shown in Fig. 8(c) and (d), respectively. The topographical histogram for S2 [Fig. 8(d)] shows a Gaussian distribution, which indicates a surface having features distributed symmetrically around a mean surface profile. The topographical histogram for S1 [Fig. 8(c)] is quite different from S2, which is an overlap of two Gaussian functions, indicating that the surface features are bimodal unlike the sputtered film. Similar behavior of surface morphology depending on growth technique was also observed earlier for Ni Film [26]. Further quantitative difference in the morphology of interfaces grown by ED and sputtering has been studied using DXRR as discussed below.

Diffuse XRR is measured at three angles of incidence which were fixed during scan and only the angle of reflection was varied around the specular peak, usually termed as detector scan [36]. Fig. 8 (e) shows the diffuse XRR as a function of in-plane momentum transfer $Q_x$, at three values of specular peak with $Q_z$ (= 0.068 Å$^{-1}$, 0.101 Å$^{-1}$, 0.148 Å$^{-1}$). The diffuse XRR data at different $Q_z$ are shifted by a factor of 10 in the plot for clarity. We analyzed diffuse XRR measurements from S1 for studying the in-plane morphology of the sample under the approximation of a self-affine fractal surface, where in-plane height-height correlation function $C(x, y)$ is usually assumed [32, 41, 42]: $C(x,y) = <\delta z(0)\delta z(x,y)> = \sigma'^2 exp\left(-\left[\frac{\sqrt{x^2+y^2}}{\xi}\right]^{2h}\right)$; Where $\sigma'$ is the *rms* value of the surface roughness, $h$ is the roughness exponent, known as Hurst parameter and $\xi$ is the in-plane correlation length of the roughness. The exponent $0<h<1$ determines the fractal dimension ($D = 3 - h$) of the interface (i.e., how jagged the interface is; smoother interfaces have larger values of $h$) [32]. For analyzing the diffuse scattering data we have used the formalism developed by Holý et al. [42] to obtain the diffuse scattering cross-section (see Eq. (16) in ref. [42]). The σ′, $h$ and ξ for each interface are the parameters of the fit to DXRR data while other parameters obtained from the specular XRR (i.e. thickness and scattering length density) were kept fixed. To increase the confidence limit for obtained parameters we have fitted DXRR data at three values of $Q_z$, with same set of parameters given in Table 3.

We obtained higher *rms* roughness for ED deposited Co layer than the other layers deposited (Cu and Ti) by sputtering (Table 3), which is also consistent with AFM measurements. DXRR results suggested a Hurst parameter, $h = 0.9$ for ED grown Co surface indicating two dimensional ($D \sim 2$) smooth growths as compared to other interfaces, which show a fractal behavior ($h = 0.35$) with dimension, $D \sim 2.6$. Inset of Fig. 8 also shows the simulated line profile of the surfaces with different Hurst parameter. We also found that the in-plane correlation length of the Co layer (ED grown) is lower ($\sim 30$ Å) as compared to other layers ($\sim 100$ Å) in the sample S1 deposited by sputtering, which indicates that heights are correlated only within the islands or intra-island (intra cluster) for the ED grown Co layer.

Thus the low density layer with specific magnetic behavior of Co film and anomalous AMR of S1 may be attributed for different morphology of ED grown film. The distinct morphology of the ED grown film may be providing excess magnetic anisotropy at the islands' edges [52] responsible for such a specific magnetic behavior. Fig. 9 shows the correlation between longitudinal magnetotransport and magnetic properties of Co films. We found very small MR for S2 as compared to that of S1. SQUID and PNR measurements also suggested similar trend for averaged in-plane magnetization of these films. Thus Co film deposited by sputtering are grown with large epitaxial strain shows low magnetization and MR. Whereas Co film deposited by ED is relaxed for epitaxial strain displays large magnetization and MR.

**Summary**

In summary we have investigated the magnetic and magnetotransport properties of Co films deposited on Si/Ti/Cu surface by two different deposition techniques: electro-deposition (ED) and sputtering. Crystalline growth of both the Co films on Si/Ti/Cu surface has been confirmed by X-ray diffraction. ED grown sample are less strained as compared to the Co film grown by sputtering. We found different growth process resulting into different morphology of the films. We observed large anisotropic magnetoresistance for electrodeposited Co film as compared to that of Co film grown by sputtering. Depth dependent structure of ED grown Co film, as obtained from XRR and PNR measurements, suggested two layer structures for Co Film with top layer (~ 100 Å) having reduced density (~ 68% of bulk Co). This two layer structure of Co film grown by ED is consistent with the histogram profile of the film measured by AFM. PNR results further revealed that the two layers (low density surface layer and rest of Co layer) are antiferromagnetically coupled. Our results indicate that the different growth and nucleation of ED grown film resulted into distinct magnetic anisotropy behavior probably caused by the islands' edges. We found excellent agreement between average magnetization of both the samples (ED and sputtered grown) measured by macroscopic SQUID and PNR. The experimental finding of very specific magnetic behavior along the depth of the Co film grown by ED will provide added interaction at surface to observe properties like magnetization reversal, magnetization controlled by electric field in Co film, which will open excellent opportunities for materials engineering and spintronics devices.

**Acknowledgement:**

We acknowledge the help of V B Jayakrishnan for XRR and XRD experiments.

Table 1: Structural parameters of Co films deposited by electro-deposition (sample S1) and sputtering (sample S2) techniques extracted from specular X-ray reflectivity measurements.

|  | Sample S1 | | | Sample S2 | | |
| --- | --- | --- | --- | --- | --- | --- |
|  | Thickness (Å) | ESLD ($10^{-5}$ Å$^{-2}$) | Roughness (Å) | Thickness (Å) | ESLD ($10^{-5}$ Å$^{-2}$) | Roughness (Å) |
| Low density Co | 105±8 | 4.6±0.20 | 25±5 |  |  |  |
| High density Co | 245±10 | 6.58±0.20 | 23±4 | 365±10 | 5.7±0.50 | 10±4 |
| Cu | 220±12 | 6.00±0.35 | 8±2 | 230±14 | 6.26±0.30 | 12±4 |
| Ti | 140±10 | 3.50±0.25 | 11±4 | 100±8 | 3.65±0.20 | 9±3 |
| Substrate | - | 2.06±0.20 | 5±2 | - | 2.06±0.20 | 5±2 |

Table 2: Structural parameters of Co films deposited by electro-deposition (sample S1) and sputtering (sample S2) techniques extracted from specular polarized neutron reflectivity measurements.

|  | Sample S1 | | | Sample S2 | | |
| --- | --- | --- | --- | --- | --- | --- |
|  | Thickness (Å) | NSLD ($10^{-6}$ Å$^{-2}$) | Roughness (Å) | Thickness (Å) | NSLD ($10^{-6}$ Å$^{-2}$) | Roughness (Å) |
| Low density Co | 105±8 | 1.6±0.25 | 18±4 |  |  |  |
| High density Co | 240±10 | 2.34±0.20 | 16±4 | 325±15 | 2.05±0.40 | 8±3 |
| Cu | 230±12 | 6.22±0.15 | 12±4 | 250±14 | 6.36±0.30 | 9±3 |
| Ti | 145±10 | -1.80±0.25 | 10±4 | 110±8 | -1.35±0.20 | 15±3 |
| Substrate | - | 2.10±0.20 | 5±2 | - | 2.10±0.20 | 6±2 |

Table 3: Morphological parameters of different interfaces of sample S1 (electrodeposited) as obtained from diffuse (Off- specular) x-ray reflectivity measurements.

|  | Sample S1 | | |
|---|---|---|---|
|  | Atomic roughness, $\sigma$ (Å) | In-plane correlation length, $\xi$ (Å) | Hurst parameter, $h$ |
| Co | 19±4 | 30±10 | 0.9±0.1 |
| Cu | 8±2 | 100±20 | 0.35±0.05 |
| Ti | 8±2 | 100±20 | 0.35±0.05 |
| Substrate | 4±2 | 100±20 | 0.35±0.05 |

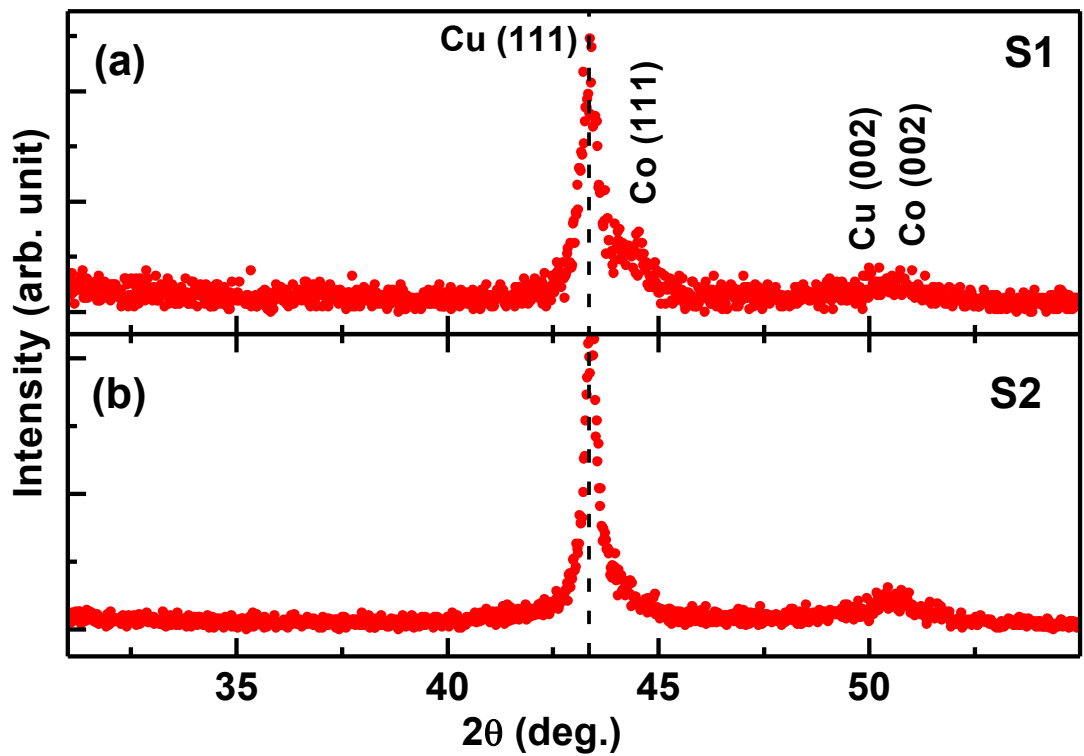

**Fig. 1:** X-ray diffraction from the Co films deposited on Si(substrate)/Ti/Cu surface using (a) electrodeposition (sample S1) and (b) sputtering (sample S2) techniques.

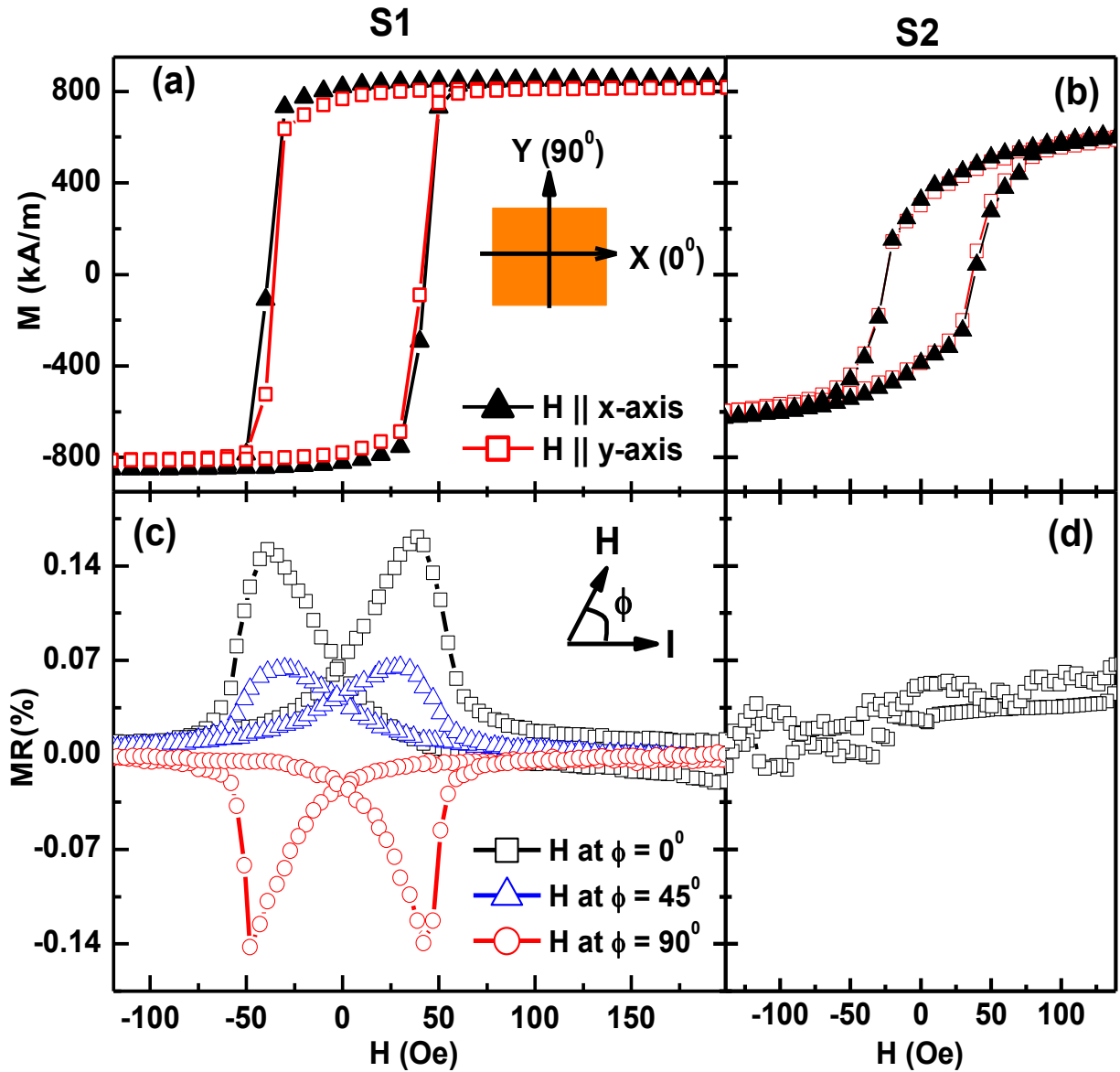

**Fig. 2:** (a) and (b) depicts the magnetic hysteresis curves, *M* (*H*), at room temperature of the Co films deposited by electrodeposition (S1) and sputtering (S2) techniques. Inset of (a) shows the schematic of two in-plane perpendicular direction along which the *M* (*H*) curves were measured. (c) Magnetoresistance (MR) data as a function of applied field from S1 in different directions, longitudinal ($\varphi = 0^0$), transverse ($\varphi = 90^0$) and at $\varphi = 45^0$, of applied field and measure current. (d) show the MR data from S2 when both applied field and current direction are parallel to each other (longitudinal).

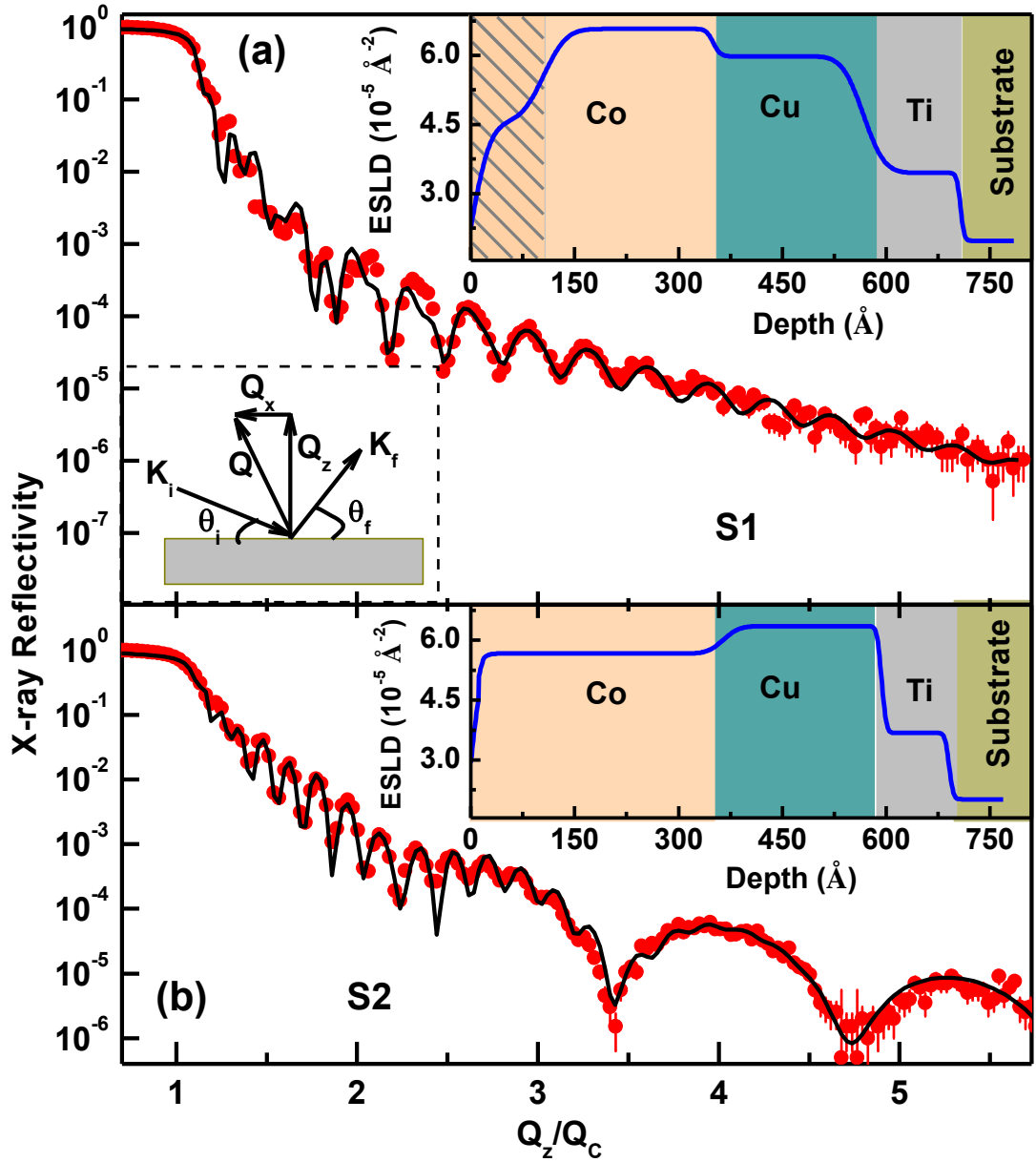

**Fig. 3:** Specular X-ray reflectivity (XRR) measurements from S1 (a) and S2 (b). Lower left inset of (a) shows the scattering geometry in reciprocal space. Insets of (a) and (b) show the corresponding electron scattering length density (ESLD) profiles of S1 and S2 which best fitted the XRR data (continuous lines in (a) and (b)).

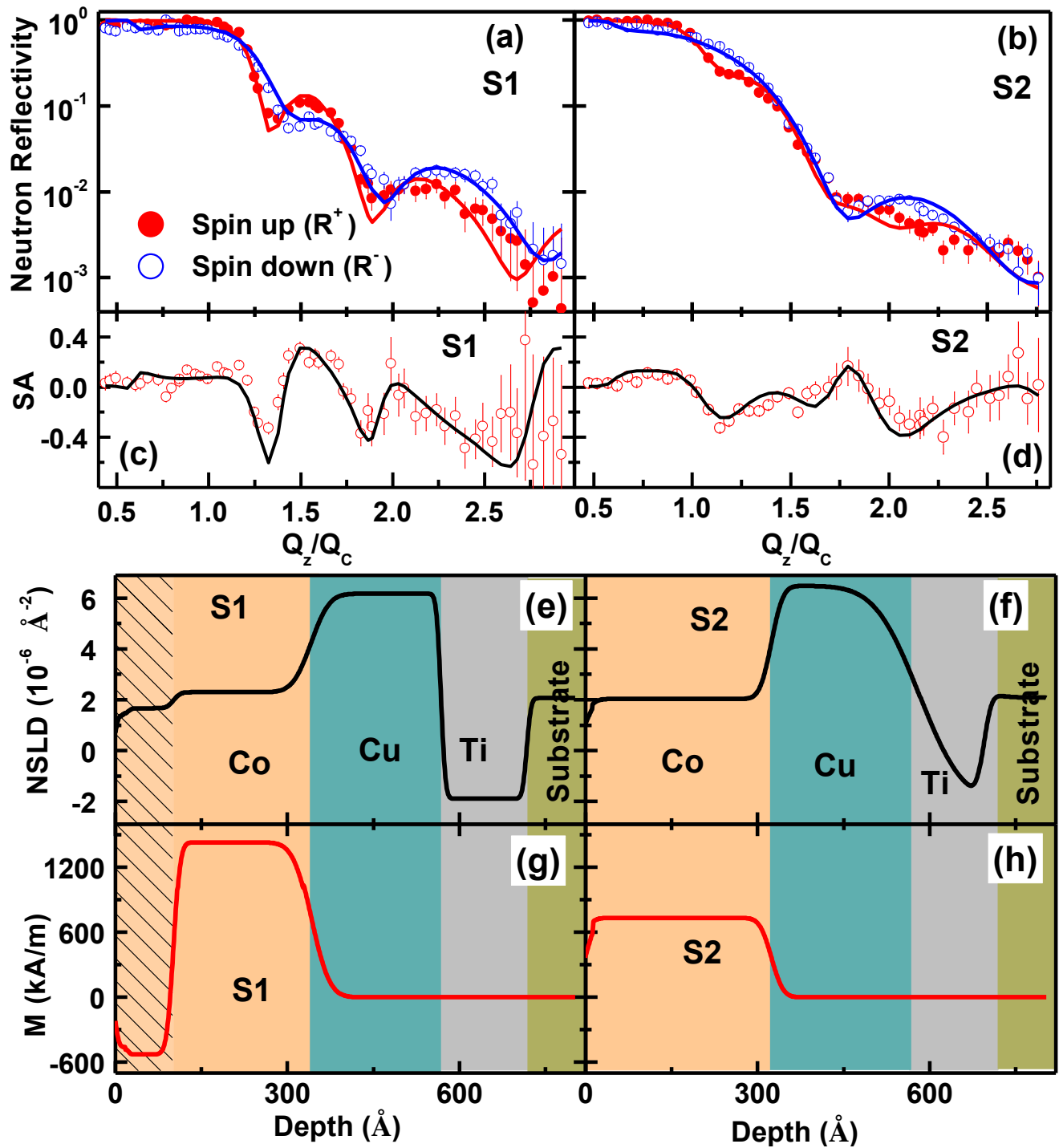

**Fig. 4:** Polarized neutron reflectivity (PNR) measurements from S1 (a) and S2 (b) at room temperature. (c) and (d) show spin asymmetry (SA) = $(R^+ - R^-)/(R^+ + R^-)$, from S1 and S2, respectively, as a function of $Q_z/Q_c$. (e) and (f) show the nuclear scattering length density (NSLD) of S1 and S2. (g) and (h) show magnetization (M) depth profile of S1 and S2. NSLD and M profiles in (e)-(h) best fitted the PNR data (continuous lines in (a) - (d)).

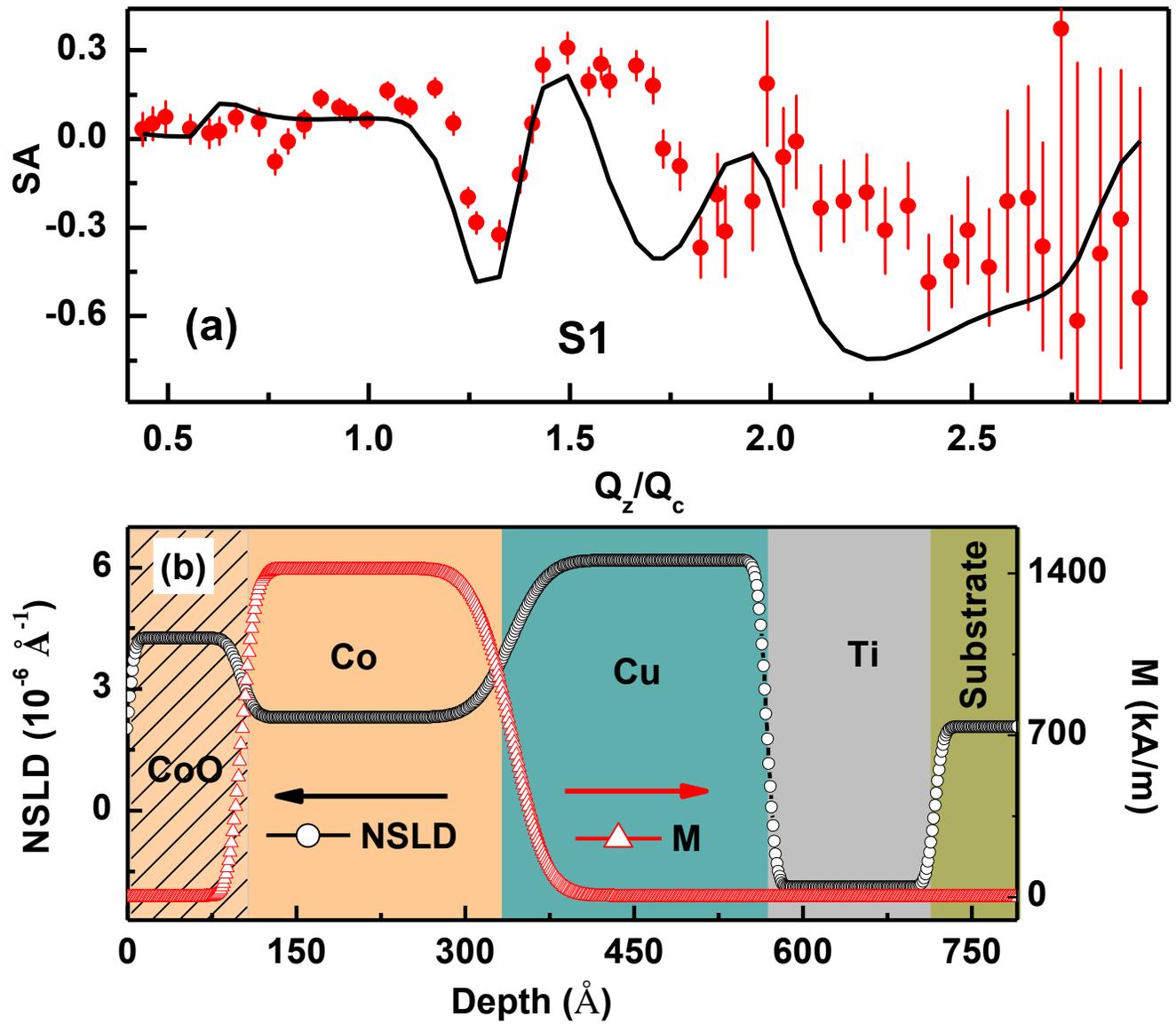

**Fig. 5:** Spin asymmetry (SA) data from S1 (a) and corresponding fit (solid line in (a)) assuming top 100 Å Co layer as CoO. (b) shows the corresponding NSLD and magnetization depth profiles.

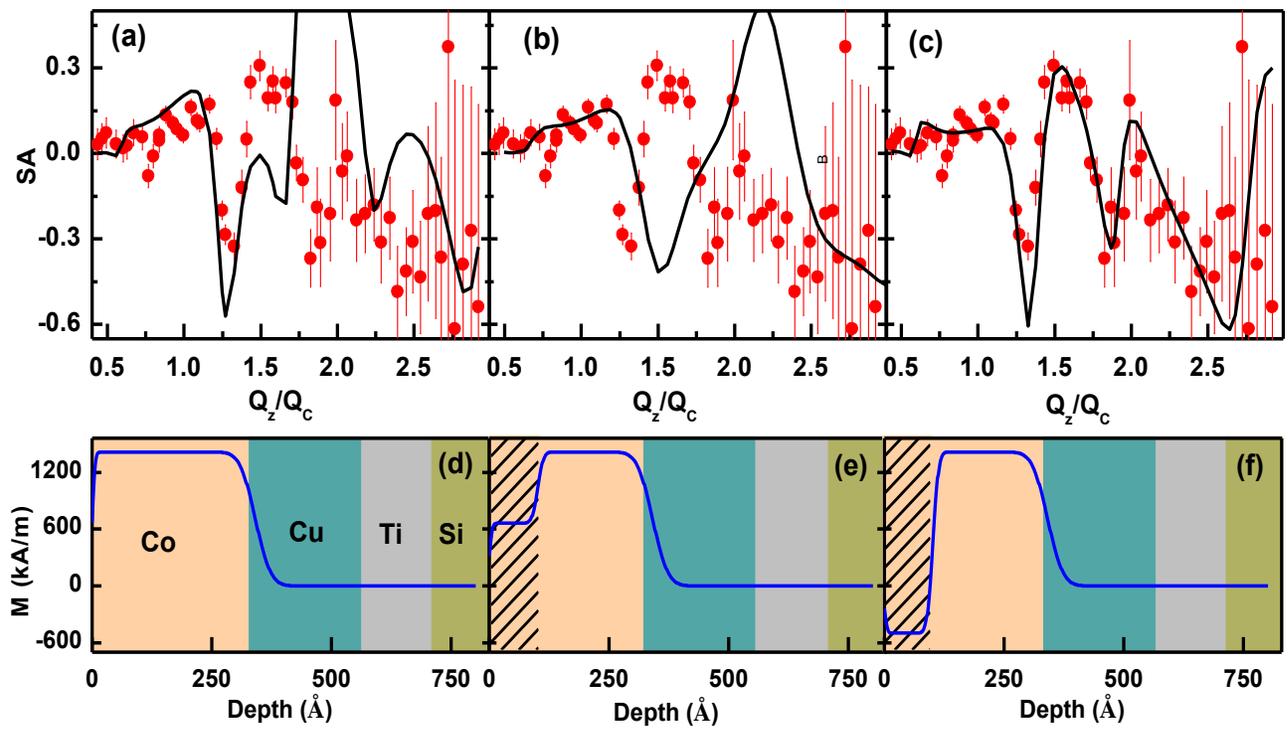

**Fig. 6:** Spin asymmetry (SA) data from S1 and corresponding fits (a-c) assuming different magnetization depth profiles (d-f) across whole Co layer in S1.

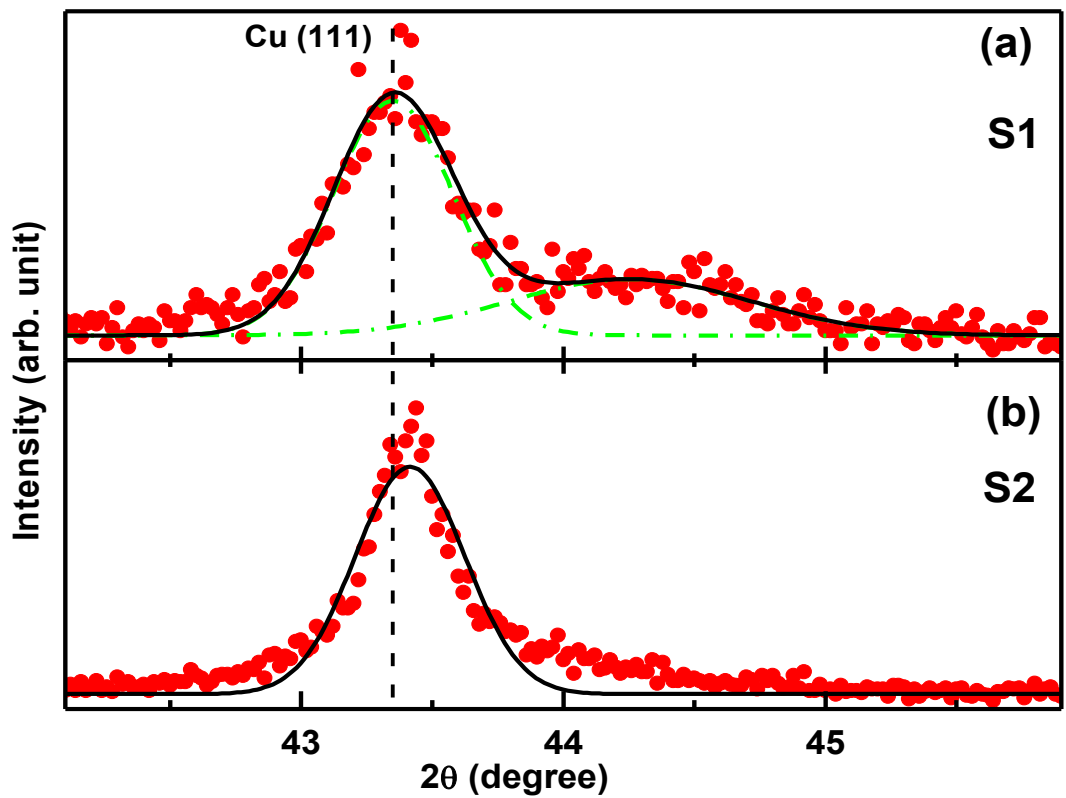

**Fig. 7:** XRD data around (111) reflection of Cu and corresponding Gaussian fit for S1 (a) and S2 (b).

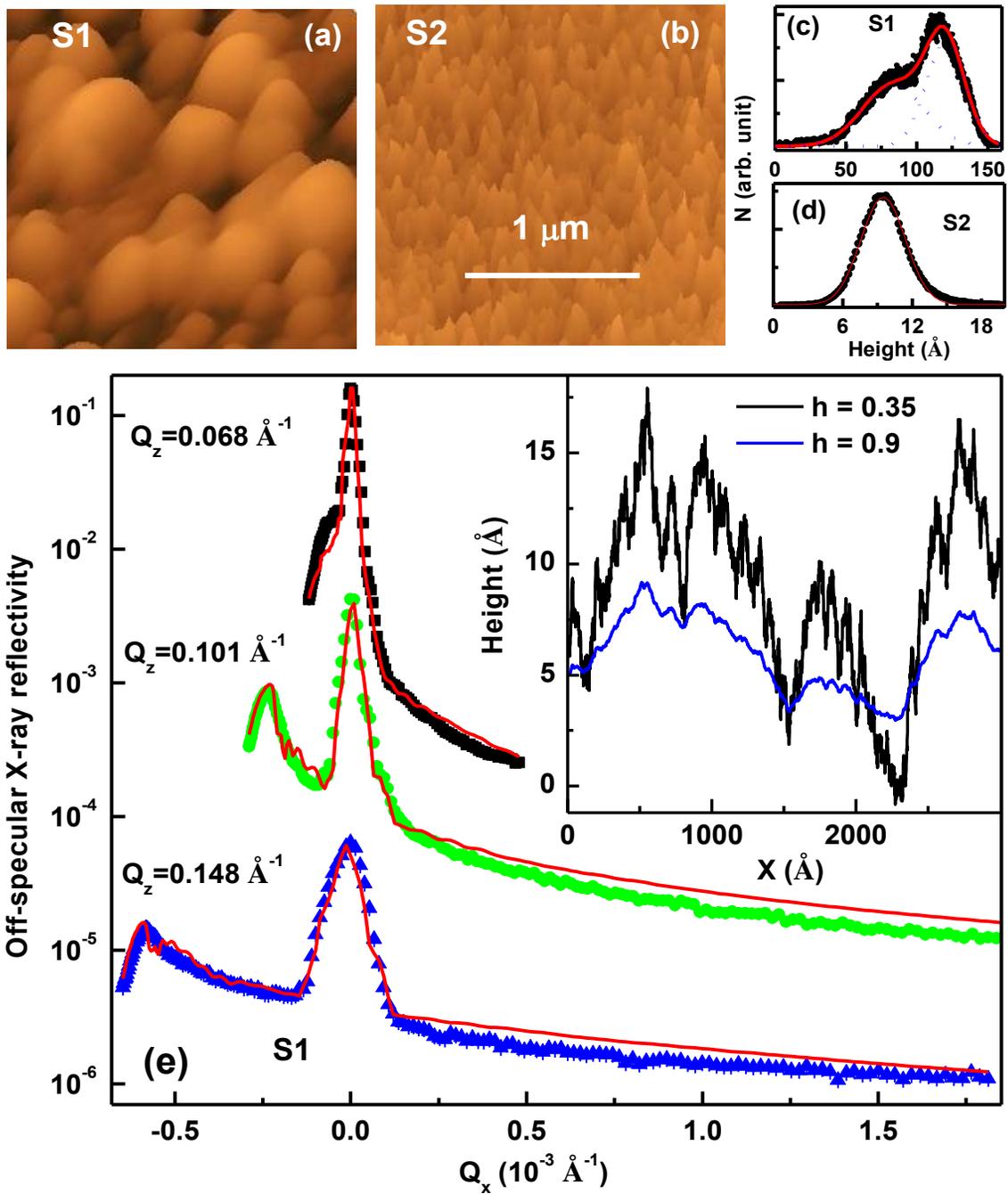

**Fig. 8:** (a) and (b) show the AFM image of the Co surface (2 μm × 2 μm) of S1 and S2, respectively. (c) and (d) show the corresponding histogram from AFM images of S1 and S2, respectively. **(e)** Off-specular X-ray reflectivity measurements from S1 at different angle of incidence (or $Q_z$). Inset show the height profile of a surface with different value of Hurst parameters (*h*).

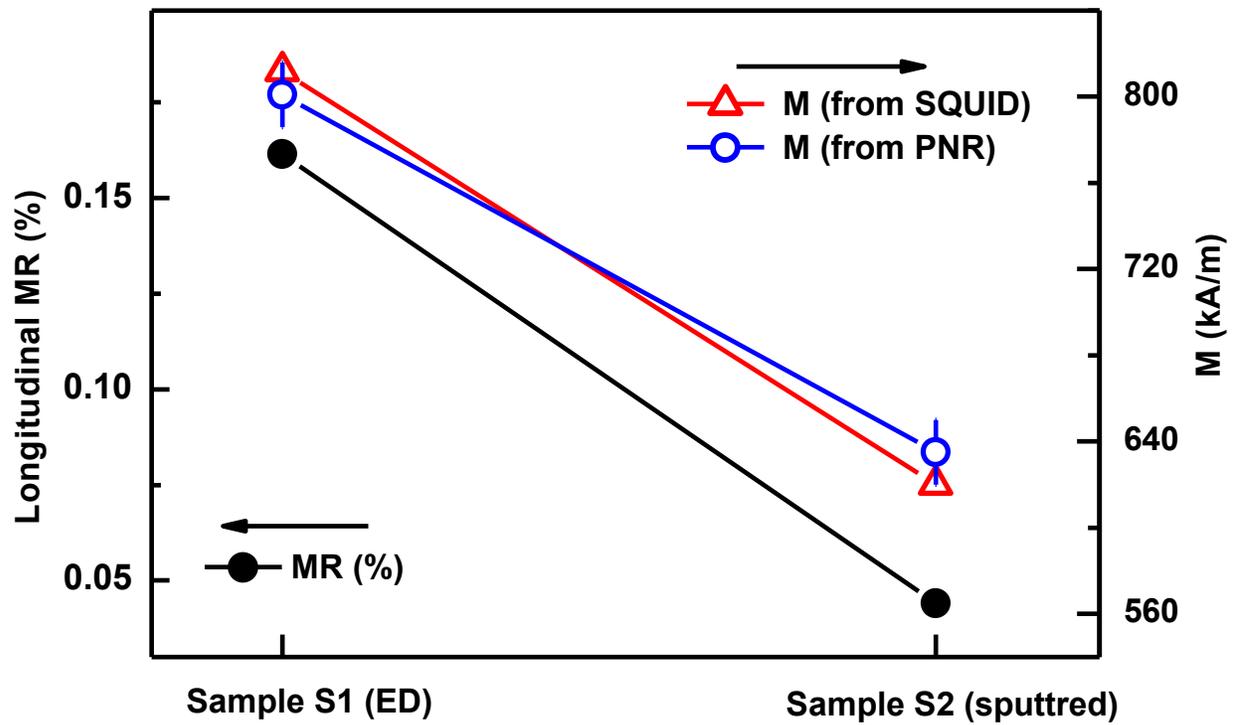

**Fig. 9:** Variation of MR and magnetization for S1 and S2. Straight line is just a guide to eye.